\documentclass[rnote]{aa} 
\usepackage{graphicx}
\usepackage{color}
\usepackage{txfonts}
\usepackage{epsfig}
\usepackage{float}
\usepackage{amsmath}
\usepackage{natbib}
\usepackage{float}
\usepackage{amssymb}
\usepackage{threeparttable}

\begin{document}
\title{Possible hard X-ray shortages in bursts from KS 1731-260 and 4U 1705-44}

\authorrunning{Ji et al.}
\titlerunning {Possible hard X-ray shortages in bursts from KS 1731-260 and 4U 1705-44}

\author{L. Ji\inst{1}\fnmsep\thanks{E-mail:  jilong@ihep.ac.cn (LJ)}, S. Zhang\fnmsep\thanks{E-mail: szhang@ihep.ac.cn (SZ)}\inst{1}, Y.-P. Chen\inst{1}, S.-N. Zhang\inst{1}, P. Kretschmar\inst{2}, J.-M. Wang\inst{1,3}, J. Li\inst{1} }
\institute{
$^{1}$Laboratory for Particle Astrophysics, Institute of High Energy Physics, Beijing 100049, China\\
$^{2}$European Space Astronomy Centre (ESA/ESAC), Science Operations Department, Villanueva de la Ca\~nada (Madrid), Spain\\
$^{3}$Theoretical Physics Center for Science Facilities (TPCSF), CAS, 19(B) Yuquan Road, Beijing 100049, China\\}

\date{Received XXX, XXXX; accepted XXX, XXXX}
\abstract
{}
{A hard X-ray shortage, implying the cooling of the corona, was observed during bursts of IGR J17473-272, 4U 1636-536, Aql X-1, and GS 1826-238. Apart from these four sources, we investigate here an atoll sample, in which the number of bursts for each source is larger than 5, to explore the possible additional hard X-ray shortage during {\it Rossi X-ray timing explorer (RXTE)} era.}
{According to the source catalog that shows type-I bursts, we analyzed all the available pointing observations of these sources carried out by the {\it RXTE} proportional counter array (PCA). We grouped and combined the bursts according to their outburst states and searched for the possible hard X-ray shortage while bursting.}
{We found that the island states of KS 1731-260 and 4U 1705-44 show a hard X-ray shortage at significant levels of 4.5 and 4.7 $\sigma$ and a systematic time lag of $0.9 \pm 2.1$ s and $2.5 \pm 2.0$ s with respect to the soft X-rays, respectively. While in their banana branches and other sources, we did not find any consistent shortage. }
{}

\keywords{stars: coronae -- stars: neutron -- X-rays: binaries --X-rays: bursts}
\maketitle

\section{Introduction}
Disc-accreting X-ray binaries (abbreviated as XRBs hereafter) are known to display distinct spectral states according to the variation of their accretion rate during the evolution of an outburst \citep{Lewin1993,Remillard2006,Esin1997}. At the high state, the spectrum is usually dominated by thermal components, and at the low state, significant non-thermal hard X-rays can be observed, which is thought to originate from the corona or the jet.
Although XRBs have been discovered decades ago, the corona formation process is still unclear. In theory, it can result from evaporation \citep{Meyer1994,Esin1997,Liu2007,Frank2002} or magnetic re-connection \citep{Zhang2000,Mayer2007}.

A hard X-ray shortage, implying the cooling of the corona, was observed during bursts of IGR J17473-272, 4U 1636-536, Aql X-1, and GS 1826-238 \citep{Maccarone2003,Chen2012,Ji2013,Chen2013,Ji2014}. Apart from these four sources, we investigate an atoll sample here (Table 1, 2), in which the number of bursts for each source is larger than 5, to explore the possible additional hard X-ray shortage during the {\it Rossi X-ray timing explorer (RXTE)}  era. In this sample, we pay extra attention to the sources that show complete color-color diagrams (CCDs; see Table 1). According to the CCDs, we can identify outburst states at a time around the occurrence of each burst and categorize the bursts into hard and soft states, respectively. Finally, we find two atoll sources  (4U 1705-44 and KS 1731-260) that showed corona cooling while bursting. In this paper, we focus on the results derived on these two sources.

The neutron star KS 1731-260 is a transient located near the Galactic center (l=1.07$^\circ$, b=3.66$^\circ$) with a mass less than 2.1 $M_\odot$ and a radius R $\leqslant$ 12.5 km \citep{Ozel2012}. It was discovered in 1989 \citep{Syunyaev1990} and remained active until transitioning to quiescence in early 2001. The distance was estimated to be $\sim$7 kpc \citep{Muno2000}.
The neutron star 4U 1705-44 is a persistently bright XRB in the direction of the Galactic bulge. \citet{Christian1997} derived a distance of 11 kpc from the peak flux of photospheric radius expansion (PRE) bursts. The kilohertz quasi-periodic oscillations were discovered using observations with {\it RXTE} by \citet{Ford1998}. The spectral analysis and fluorescent iron line have been carried out by \citet{DiSalvo2005}.

Section 2 introduces the data and the observations we analyze. Section 3 describes the results of the corona cooling and their intrinsic timescale of reheating.The summary and discussion of these results are provided in Section 4.

\begin{table*}
\tiny
\begin{center}
\caption{The atoll sources that show complete CCDs and have a burst number larger than 5. The columns denote the source name, number of bursts, averaged persistent flux at 40-50 keV and 3.6-18 keV, and peak flux of bursts at 3.6-18 keV for hard and soft states.}
\begin{tabular}{|c|c|c|c|c|c|c|c|c|}
\hline
&\multicolumn{4}{|c|}{hard state} & \multicolumn{4}{|c|}{soft state} \\
\hline
source        & number    &   persistent flux at &  persistent flux at  &    peak flux at             &    number         &    persistent flux at      &  persistent flux at  &    peak flux at             \\
              &           &    40-50 keV (cts/s) &  3.6-18 keV (cts/s)  &  3.6-18 keV (cts/s)         &                   &     40-50 keV (cts/s)      &  3.6-18 keV (cts/s)  &  3.6-18 keV (cts/s)         \\
\hline
4U 1608-52    &    21     &   0.220 $\pm$ 0.018  &   106.67  $\pm$ 0.07   &  7065.73  $\pm$  24.88    &     12            &     0.011 $\pm$ 0.036      &  436.73  $\pm$  0.13      &  8760.11  $\pm$  40.66            \\
4U 1702-429   &    6      &   0.155 $\pm$ 0.011  &   90.47   $\pm$ 0.05   &  5253.94  $\pm$  33.79    &     3             &    -0.042 $\pm$ 0.006      &  131.64  $\pm$  0.03      &  3539.94  $\pm$  11.05            \\
4U 1728-34    &    39     &   0.245 $\pm$ 0.007  &   218.72  $\pm$ 0.05   &  6123.01  $\pm$  18.98    &     17            &     0.033 $\pm$ 0.011      &  342.89  $\pm$  0.09      &  4204.80  $\pm$  19.96            \\
EXO 0748-676  &    77     &  -0.003 $\pm$ 0.005  &    21.09  $\pm$ 0.01   &  856.74   $\pm$  3.69     &     9             &    -0.058 $\pm$ 0.015      &  21.16   $\pm$  0.04      &  1164.84  $\pm$  13.12            \\
KS 1731-260   &    16     &   0.337 $\pm$ 0.008  &   126.21  $\pm$ 0.05   &  2377.92  $\pm$  12.48    &     1             &    -0.142 $\pm$ 0.023      &  455.30  $\pm$  0.22      &  3792.18  $\pm$  68.01            \\
4U 1705-44    &    30     &   0.247 $\pm$ 0.008  &    93.71  $\pm$ 0.04   &  1486.46  $\pm$  7.71     &     14            &     0.003 $\pm$ 0.012      &  124.57  $\pm$  0.06      &  2099.39  $\pm$  12.49        \\
\hline
\end{tabular}
\end{center}
\end{table*}

\begin{table*}\tiny
\begin{center}
\caption{The sources that have a burst number larger than 5 but show no complete CCDs. The columns denote the source name, number of bursts, averaged persistent flux at 40-50 keV and 3.6-18 keV, and peak flux of bursts at 3.6-18 keV.}
\scalebox{1.0}{
\begin{tabular}{|c|c|c|c|c|}
\hline
source        & number   &  persistent flux at  &    persistent flux at  &   peak flux at          \\
              &          &   40-50 keV (cts/s)  &    3.6-18 keV (cts/s)  &  3.6-18 keV (cts/s)     \\
\hline
4U 0513-401      &    11               &  -0.033 $\pm$ 0.034      &     34.89       $\pm$  0.06    &  1126.33  $\pm$ 10.28  \\
x1735-444        &     8               &  -0.026 $\pm$ 0.010      &     228.46      $\pm$  0.09    &  2374.16  $\pm$ 18.75  \\
4U 1820-30       &    12               &   0.230 $\pm$ 0.009      &     227.46      $\pm$  0.06    &  3906.67  $\pm$ 19.17 \\
HETEJ1900.1-2455 &    10               &   0.203 $\pm$ 0.015      &      60.76      $\pm$  0.05    &  5635.34  $\pm$  27.54 \\
1M 0836-425      &    15               &   0.273 $\pm$ 0.011      &      79.82      $\pm$  0.04    &  1030.44  $\pm$  8.41  \\
EXO 1745-248     &    16               &   0.478 $\pm$ 0.012      &      215.18     $\pm$  0.08    &  1540.94  $\pm$ 11.48   \\
4U 1916-053      &    12               &   0.072 $\pm$ 0.009      &      27.45      $\pm$   0.02   &   2031.05  $\pm$  13.24\\
IGR J17511-3057  &     9               &   0.234 $\pm$ 0.008      &      45.88      $\pm$   0.03   &   2945.85  $\pm$  18.60 \\
SLX 1744-300     &     8               &   0.089 $\pm$ 0.010      &      82.29      $\pm$   0.04   &    779.84  $\pm$  10.01 \\
SAX J1750.8-2900 &     6               &   0.109 $\pm$ 0.019      &      126.83     $\pm$   0.09   &    2965.07 $\pm$  22.84 \\
XTE J1759-220    &     6               &   0.045 $\pm$ 0.011      &      25.47      $\pm$   0.03   &    695.87  $\pm$  10.90  \\
\hline
\end{tabular}
}
\end{center}
\end{table*}

\section{Observation and data reduction}

According to the source catalog that shows type-I bursts and is provided by \citet{Galloway2008}, we analyzed all the available pointing observations of these sources carried out by {\it RXTE} proportional counter array (PCA). Apart from those four for which the hard X-ray shortages were reported while bursting, we select the sources in which larger than 5 bursts were observed for the following analysis and present them in Table 1 and Table 2. In Table 1, we have six atolls for which the complete CCDs are derived. In Table 2, we have eleven atolls that only parts of the CCDs are available from the observational data.
Among the five co-aligned Xe multiwire proportional counter units (PCUs), only PCU2 was used due to its better calibration and longer coverage.  We used the data of the Standard 2 mode when producing CCDs, while only the data from the E\_125u\_64M\_1\_s and the Goodxenon mode were adopted when studying the properties of bursts due to their high resolution. The data were filtered with the standard criteria: the elevation angle is larger than 10$^{\circ}$, and the pointing offset is less than 0.02$^{\circ}$.
The background files were created using the program {\it pcabackest} with the latest bright source background model,  and the detector breakdowns have been removed.
We should note that PCA count rates are dominated by the background at 40-50 keV. Therefore, the statistical fluctuation of the background could reduce the significance level for the possible hard X-ray shortages. However, the systemic changes of background, which depend on orbital position of the satellite, would give small values because of the short duration of bursts \citep{Jahoda1996}.
The dead time correction was made under the canonical criterion described at the HEASARC Web site \footnote{Please see the website:\\ http://heasarc.nasa.gov/docs/xte/recipes/pca\_deadtime.html} with the standard 1b data as a reference.
The data analysis was performed with {\it HEAsoft ver 6.12}. The spectra were fitted with {\it XSPEC 12.7.1}, and an additional 1\% systematic error was added because of the calibration uncertainties.

\begin{figure}
\centering
\includegraphics[width=2.5in]{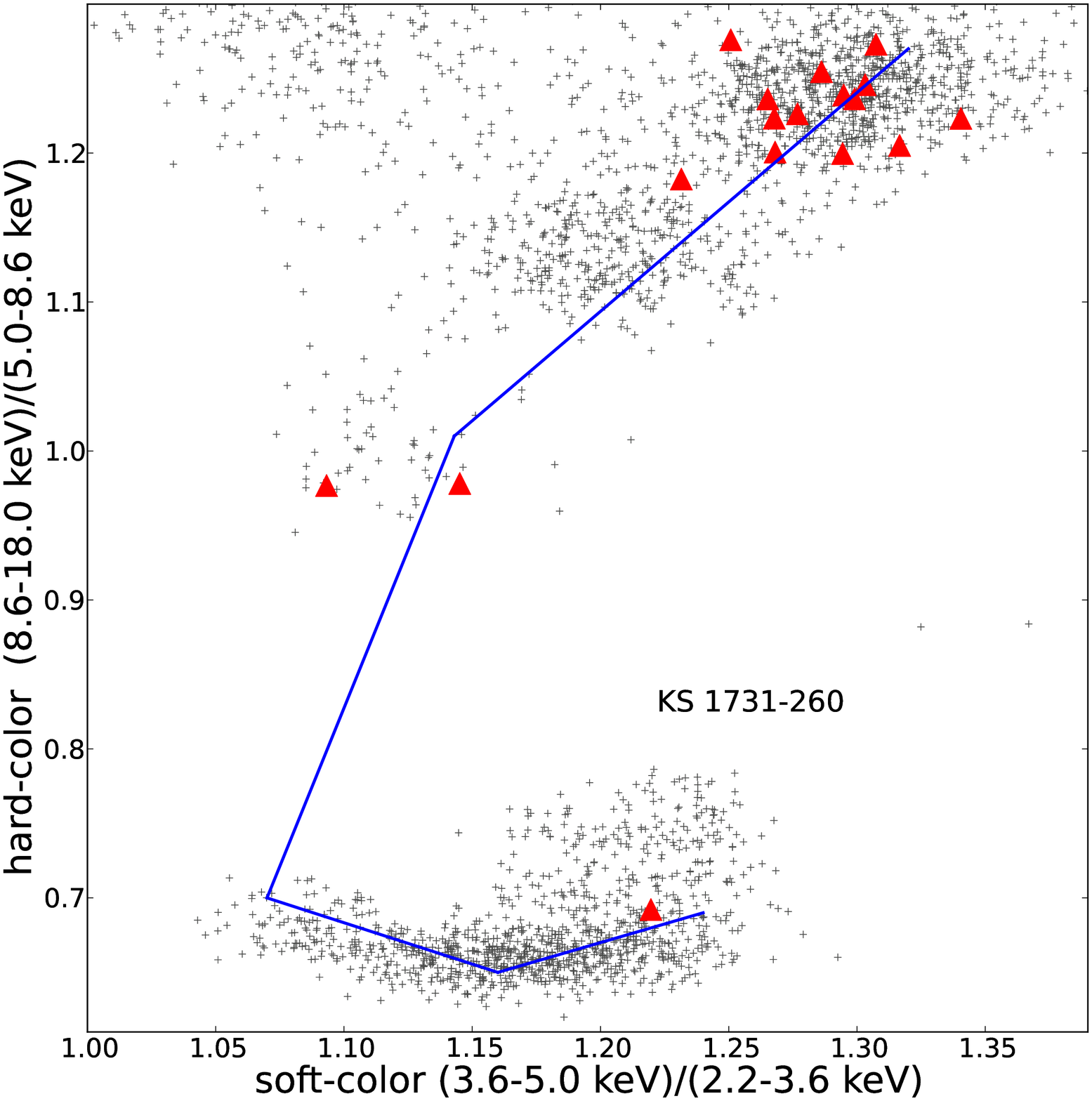}
\includegraphics[width=2.5in]{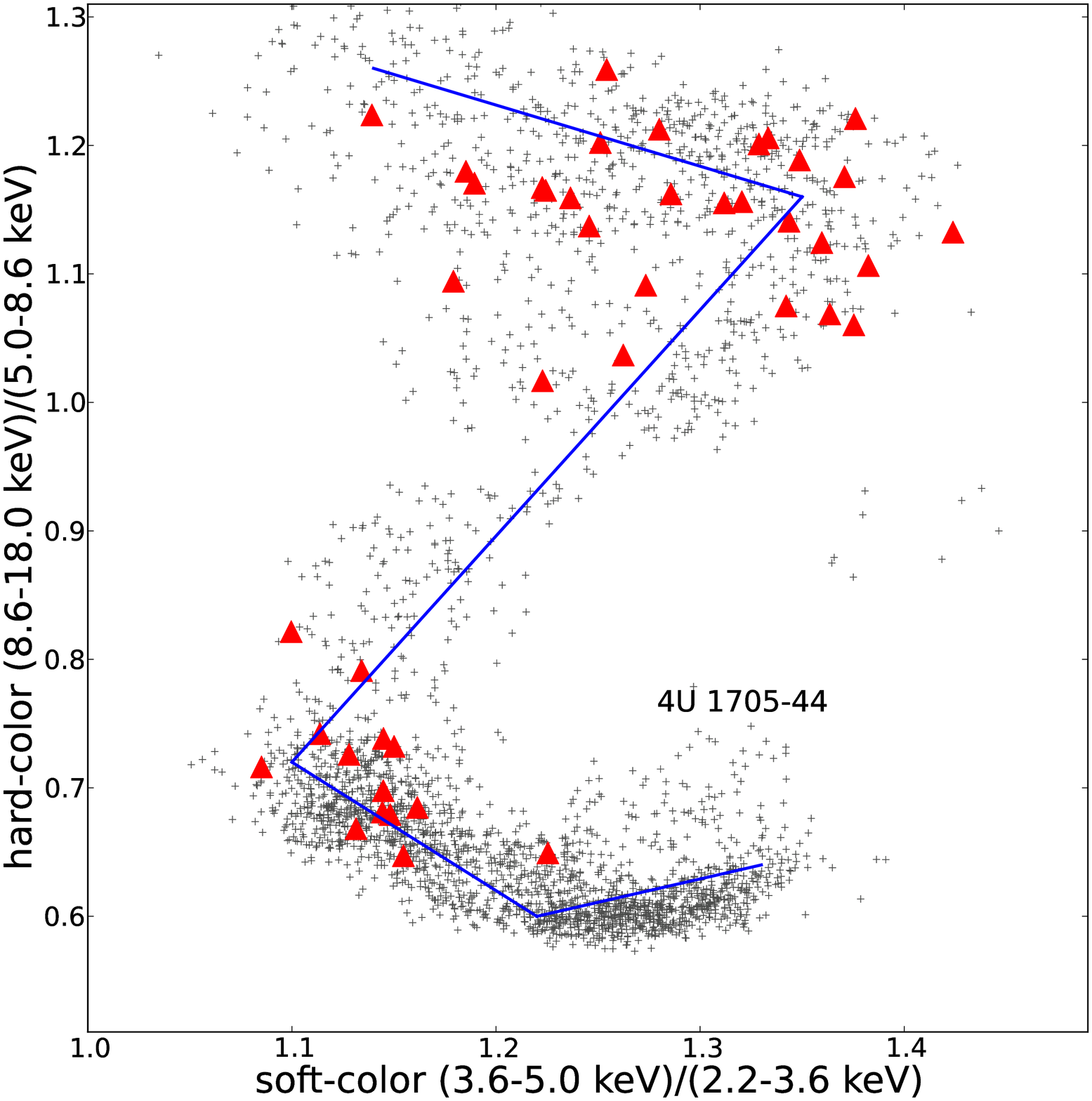}
\caption{CCDs for KS 1731-260 and 4U 1705-44, which define a soft color as a ratio of 3.6-5.0 keV to 2.2-3.6 keV and a hard color as a 8.6-18.0 keV over 5.0-8.6 keV ratio with the bin size of 128s. The bursts are located with red triangles. The blue lines sketch their "atoll" tracks.}
\label{colorcolor}
\end{figure}

To study the outburst evolution of the sources on the CCDs, we defined the soft and hard colors as the ratios of the background subtracted counts at energies of 3.6 -- 5.0 keV to  2.2 -- 3.6 keV and  of 8.6 -- 18.0 keV to  5.0 -- 8.6 keV, respectively. We then defined broken lines to represent the atoll shapes. We then assigned $S_a =1 $ to the upper right vertex of CCDs and $S_a =2 $ to the lower right vertex. The distance along the broken line between the two points can be regarded as the unit length.
Due to the data used here, which spanned five different {\it RXTE} gain epochs, and the slow drift of the PCA gain over time because of the change of the voltage settings, we analyzed the data of Crab to get the correct channel-to-energy conversions by assuming that the intensity of the Crab Nebula with PCU2 was constant \citep{Belloni2010}. Namely, we normalized each point by the closest Crab observation that was enclosed in  the same gain epoch.

We excluded bursts for which the peak flux are not 300 cts/s larger than the persistent emission.
In 4U 1636-536 and Gs 1826-238, an energy band of 30-50 keV is used to study the properties of the corona because the temperature of type-I bursts is relatively low in these two bursters\citep{Ji2013,Ji2014}.
However, the burst temperature in some sources (e.g. 4U 1728-34) in Table 1 and 2, is much higher.
Therefore, we took a narrower energy band of 40-50 keV to indicate the intensity of the corona to minimise the contaminations by the bursts, following \citet{Chen2013}.
Using the date of Standard 2 mode, the persistent flux for each burst is estimated as the average count rate of the whole observation ID at 40-50 keV during which the data at $\sim$ 100 seconds around the burst is subtracted off.
The persistent flux at energies of 3.6-18 keV and 40-50 keV shown in Table 1 and 2 represents the averaged persistent flux for the bursts in a source or a subgroup.
Following a standard approach, we took the flux at a window of 30s prior to each burst as the background and subtracted it off to have the net lightcurve and spectrum \citep{vanParadijs1986,Lewin1993}.
The cross correlation was calculated by {\it CROSSCOR} \footnote{The cross correlation was calculated with the software CROSSCRO, and detailed information is available at http://heasarc.gsfc.nasa.gov/ftools/fhelp/crosscor.txt}, a  standard software in {\it XRONOS}, which computes the coefficient normalized by the square root of the number of newbins with the fast fourier transform (FFT) algorithm.

\section{Results}

\begin{figure*}
  \centering
  \includegraphics[width=2.5in]{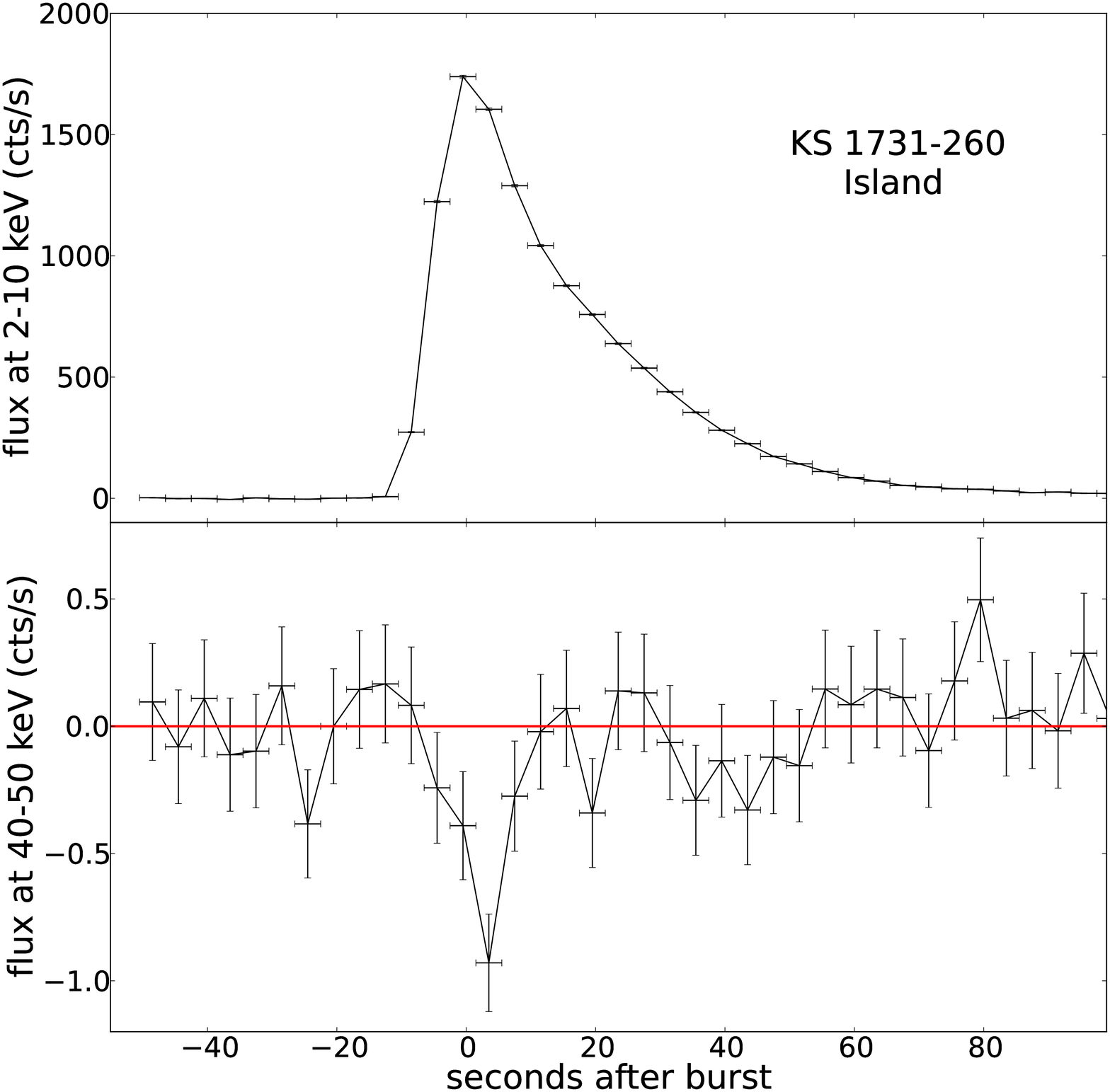}
  \includegraphics[width=2.5in]{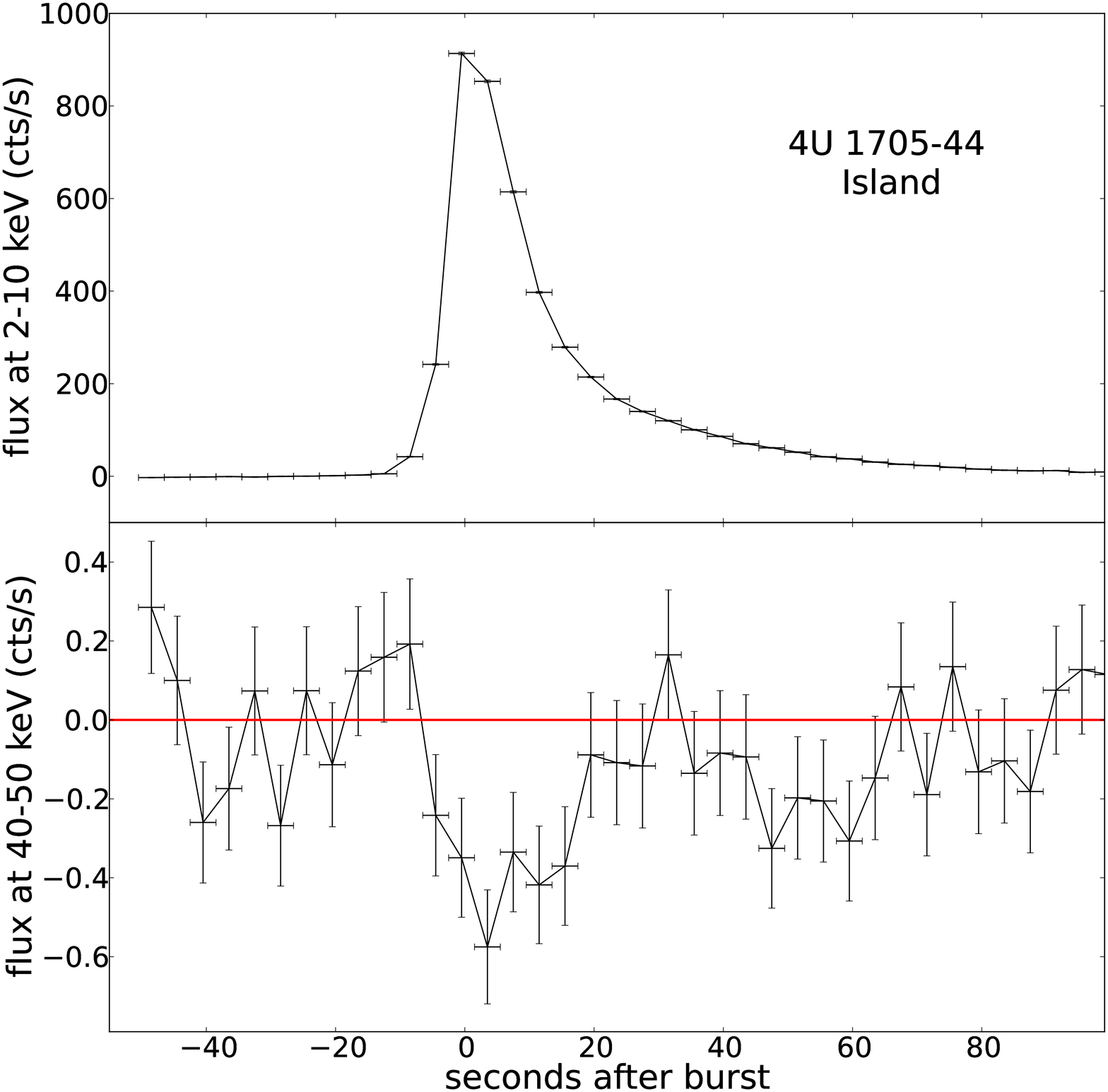}
  \caption{The combined 4 s-bin burst lightcurves for island branches of KS 1731-260 and 4U 1705-44, as derived at the 2--10 keV and 40--50 keV bands, respectively.}
\label{shortage}
\end{figure*}

The CCDs of the sources in Table 1 are produced with a bin size of 128 s and the bursts are marked with red triangles (see Figure~\ref{colorcolor} for example). The value of $S_a$ for each burst is calculated by finding the nearest point on the blue curve. We define the island (banana) branch with $S_a$ smaller (larger) than 1.8. Therefore, we classify the bursts into two subgroups for each source: the island group and banana group.
For each group, we combine the 2--10 keV and 40--50 keV lightcurves respectively and average them in each time bin by taking the 2--10 keV flux peak as a reference for each burst.
We find that the bursts induce hard X-ray shortage at significance levels of 4.5 and 4.7 $\sigma$ in the hard states of 4U 1705-44 and KS 1731-260, respectively (shown in Figure~\ref{shortage}).
The significant level is estimated by ${\chi}^2$ test with the null hypothesis that the lightcurve is a constant line.
While neither in their banana branches nor for 4U 1608-52, 4U 1728-34, 4U 1702-429, and EXO 0748-676 there shows no effect of shortage at hard X-rays.
For the sources in Table 2, we simply combined all the bursts, since there are no complete CCDs for inferring their spectral states. However, no source shows the hard X-ray shortage.
In the two sources, we find that the hard X-ray shortages are comparable with their corresponding persistent flux, which may imply most of the corona is cooled off.

Regarding to the bursts born in these two sources, we perform a cross correlation study of the 8-s bin \footnote{The different bin size has little effect on the time delay, but the smaller bin size leads to a smaller value of cross correlation due to the statistical fluctuation .
With respect to the following simulation, the mean cross correlation coefficients are -0.55 and -0.51, respectively.}
lightcurves at energies between 2--10 keV and 40--50 keV. The time delays and their errors are calculated by a throwing-dot method.
In practice, we sample the lightcurves at energies between 2--10 keV and 40--50 keV by assuming that each point in lightcurves has a Gaussian distribution and then calculate the time delay with a cross-correlation method. By sampling the lightcurve a thousand times, we fit the resulted time delay and corresponding error with a Gaussian function. The resulted time delays in KS 1731-260 and 4U 1705-44 are $0.9 \pm 2.1$ and $2.5 \pm 2.0$, respectively.

\section{Summary and discussion}
We report on the hard X-ray shortage during bursts in KS 1731-260 and 4U 1705-44, suggesting the corona can be cooled by the shower of seed photons from type-I bursts.  For all the six sources (the previous four and the two reported in this paper) so far showing corona cooling, the reheating timescale is $\sim$ seconds. However, a similar phenomenon cannot be observed for other sources in Table 1 and 2.

\subsection{Comparison between the sources showing the corona cooling}
Here, we find two more atoll sources, which behaved as possible hard X-ray shortages. These, together with the previous four sources (Aql X-1, IGR J 17473-2721, 4U1636-536, and GS 1826-238) constitute six sources exhibiting cooling of the corona during type-I bursts.
However, the evolution of the outbursts in these sources and the position in the CCDs when the bursts occurred are quite different. The bursts in IGR J17473-2721 are embedded in a long-lived preceding hard state of a single outburst \citep{Chen2012}. The accretion rate in GS 1826-238 ranges from $5\% {L}_{\rm edd}$ to $9\% {L}_{\rm edd}$, which is remarkably stable.
The luminosity of 4U 1636-536, KS 1731-260, and 4U 1705-44 is  $3\%-16\% {L}_{\rm edd}$, $6\%-38\% {L}_{\rm edd}$, and $1.1\%-70\% {L}_{\rm edd}$, respectively. The bursts in Aql X-1 are mostly located in the failed outbursts (the outbursts without state transitions). This implies that the cooling of the corona prompted by the bursts may be independent of the detailed accretion evolution.
In addition, it seems that the different duration of the bursts (for example, $\sim$ 30s in 4U 1705-44 and $\sim$ 100s in GS 1826-238), which suggest the different proportion of hydrogen in accretion materials,
do not greatly affect the reheating process because of their similar time delays.
This result seems to suggest that the physical process dominating the corona is similar in different sources.

\subsection{Cooling and reheating}
By assuming the equipartition between the thermal and magnetic energy ($ \frac{3}{2}nkT \sim  B^2/8\pi$) or the balance condition ($c_{\rm s}^2 \rho \sim B^2/8\pi$) in typical ADAF models, the magnetic field in the corona is $\sim 10^6$ G.
Given that the relation of the cooling efficiency between Compton scattering and synchrotron radiation is $\frac{P_{\rm comp}}{P_{\rm sy}} \simeq \frac{U_{\rm ph}}{U_{\rm B}}$, the corona cooling is dominated by the inverse Comptonization.
Since an effective Compton cooling of a pure electron plasma takes $\sim 10^{-6}$ s, the discovered time lag of about a few seconds should be intrinsic to the corona recovery process.

In theory, evaporation models give a timescale of days, and it is usually used to explain the behavior of the corona evolution in long-term outbursts. \citet{Ji2014} proposed that the timescale for evaporation while bursting could drop significantly after considering the additional angular momentum transfer mechanism. The detailed reheating model, however, remains unknown.

Alternative to the evaporation model, the corona can also be formed through a magnetic re-connection process, as was firstly proposed according to the similarity between the XRB disks and the solar corona. In a magnetic re-connection model, similar to what happens in Sun, magnetic turbulence and buoyancy can trigger magnetic flares and cause an intense heating  \citep{Zhang2000}.
Following the model by \citet{Liu2002}, the energy balance in the corona between heating by magnetic reconnection and cooling by Compton scattering, is as follows:
\begin{equation}
\label{ene_bal}
\frac{{{B^2}}}
{{4\pi }}{V_A} \approx \frac{{4kT}}
{{{m_e}{c^2}}}n{\sigma _T}c{U_{\rm rad}l},
\end{equation}
where $V_A \sim {B}/{\sqrt{4 \pi \rho_0 }}$ is Alfv\'{e}n speed. Here, $\rho_0$ is the electron density, $\sigma _T$ is
the Thomson cross-section , $U_{\rm rad}$ is the soft photon field to be Compton scattered,  $U_{\rm rad} \sim {\sigma T_
{\rm disc}^4}/{c} \sim 2 \times {10}^{13}$ erg {cm}$^{-3}$, $\sigma$ is the Stefan-Boltzmann constant, and $l$ is the characteristic
length of the magnetic loops.
We estimate the timescale of the reheating process in a magnetic reconnection model as $l/c_{\rm s}$, where $c_{\rm s}$ is the sound speed and $c_{\rm s} \simeq 10 (T/10^4)^{1/2}$ km s$^{-1}$ \citep{Frank2002}. By assuming the $l\sim 10\ r_g$ and a characteristic temperature between $10^7 K$ and $10^9 K$, the time delay ranges from 0.01 to 0.1s.

\subsection{The atoll sources without hard X-ray shortage}
Except for KS 1731-260 and 4U 1705-44, we find no significant hard X-ray shortage in other sources listed in both Table 1 and Table 2.
During banana states for atoll sources enclosed in Table 1, there is no evidence of hard X-ray shortage probably because the persistent hard X-ray flux itself is too weak to be detectable. This could be the case as well for some sources listed in Table 2.

Apart from  KS 1731-260 and 4U 1705-44, there are seven other sources in Tables 1 and 2 that have the 40-50 keV flux larger than 0.2 cts/s. However, we see no obvious hard X-ray shortage while bursting from these seven sources. One possibility  that makes shortages invisible could be that, bursts with high temperature contribute hard X-rays that are not negligible while bursting to dilute the original shortage. For example, the mean temperature for 4U 1728-34 around the peak of the bursts is 2.8 keV, which leads to a contribution of 0.45 cts/s at 40-50 keV.
As once discussed in Chen et al. (2012) in the case of the lacking shortage in the lagging low hard state of IGR J17473-2721, an additional possibility is that the corona may be located too far away from the neutron star to be effectively cooled off by the bursts. This occurs because in current observations the innermost radius of the accretion disk in the hard state can not be well constrained due to the relative low luminosity and severely hardened spectrum.  Another possibility could be that the hard X-rays are from the jet, which is usually hard to be cool off  under the soft X-ray shower of the bursts.

\subsection*{Acknowledgments}
We acknowledge support from 973 program 2009CB824800 and the National Natural Science Foundation of China via NSFC 11073021, 11133002, 11103020 and XTP project XDA04060604. DFT work is done in the framework of the grants AYA2012-39303, SGR2009-811, and iLINK2011-0303. DFT was additionally supported by a Friedrich Wilhelm Bessel Award of the Alexander von Humboldt Foundation. This research has made use of data obtained from the High Energy Astrophysics Science Archive Research Center (HEASARC), provided by NASA Goddard Space Flight Center.

\end{document}